# Sub-luminous type Ia supernovae from the mergers of equal-mass white dwarfs with M~0.9 M⊙

Rüdiger Pakmor, Markus Kromer, Friedrich K. Röpke, Stuart A. Sim, Ashley J. Ruiter & Wolfgang Hillebrandt

*Max-Planck-Institut für Astrophysik, Karl-Schwarzschild-Str. 1, 85748 Garching, Germany*

**Type Ia supernovae (SNe Ia) are thought to result from thermonuclear explosions of carbon-oxygen white dwarf stars[1]. Existing models[2] generally explain the observed properties, with the exception of the sub-luminous 1991-bg-like supernovae[3]. It has long been suspected that the merger of two white dwarfs could give rise to a type Ia event[4,5], but hitherto simulations have failed to produce an explosion[6,7]. Here we report a simulation of the merger of two equal-mass white dwarfs that leads to an underluminous explosion, though at the expense of requiring a single common-envelope phase, and component masses of ~0.9 M⊙. The light curve is too broad, but the synthesized spectra, red colour and low expansion velocities are all close to what is observed for sub-luminous 1991bg-like events. While mass ratios can be slightly less than one and still produce an underluminous event, the masses have to be in the range 0.83-0.9 M⊙.**

Thermonuclear burning in normal SNe Ia produces 0.4 to 0.9 M⊙[8] of $^{56}$Ni, the radioactive decay of which powers the optical display of SNe Ia and makes them amongst the most luminous objects in the Universe. This range of $^{56}$Ni masses can be reproduced by explosion models of white dwarfs which have masses at the Chandrasekhar limit[2,9]. However, 1991bg-like supernovae are characterized[3] by a B-band peak magnitude of ~ -17 mag, pointing to $^{56}$Ni masses of only about 0.1 M⊙[8]. This sub-class has to be understood in order to theoretically capture the full range of SN Ia diversity. Interestingly, a recent analysis of the spectral evolution of SN 2005bl[10], which is representative of the 1991bg-like SNe Ia, showed that both iron group elements and silicon are present over a wide range of radii extending as close to the centre of the ejecta as is accessible observationally. It is hard to imagine an explosion model in which hydrodynamic processes alone account for such a strong mixing of silicon and iron group elements as observed in SN 2005bl. In normal SNe Ia the core of the ejecta consists predominantly of iron group elements which is surrounded by a silicon-rich layer. This difference in the characteristic chemical structures points to a different burning regime realized in the 1991bg-like objects.

Incomplete silicon burning is a natural way of producing a mixture of iron group elements and silicon. It occurs in low-density carbon-oxygen fuel for a narrow window of ash temperatures[11]. Burning significant amounts of stellar material in this regime therefore requires a shallow density profile of the exploding object. Moreover, supersonic propagation of the burning front is required to avoid pre-expansion of the material. The model described below satisfies both of these conditions and yields a small $^{56}$Ni mass.

In contrast to previous merger simulations[6,7] which considered white dwarfs of significantly different masses, our model assumes equal-mass white dwarfs. Both have a central density of $1.4 \times 10^7$ g cm$^{-3}$, a mass of 0.89 M$_\odot$, a composition of equal parts by mass of carbon and oxygen and an initial temperature of $5 \times 10^5$ K. The initial orbit is circular with a period of 28s. This corresponds to the state when tidal forces deform the white dwarfs sufficiently to make the system unstable (prior evolution is driven by gravitational wave emission and is not simulated). This progenitor system is set up as the initial conditions for a simulation using a smoothed particle hydrodynamics code[12] (SPH). In the initial conditions, marginal asymmetries were deliberately introduced since perfect symmetry is not expected in Nature. In this first simulation, we follow the inspiral and subsequent merger of the binary system (Figure 1). After two orbits one of the two white dwarfs is disrupted. This unequal evolution of the white dwarfs originates from the symmetry-breaking in the initial conditions. The disrupted white dwarf violently merges with the remaining white dwarf and material is heated by compression. In the hottest regions carbon burning begins and releases additional energy which further heats the material. A hotspot, which is resolved by several SPH particles, forms with a temperature of $2.9 \times 10^9$ K in high-density material ($3.8 \times 10^6$ g cm$^{-3}$). High-resolution small-scale simulations[13] show that under such conditions a detonation ignites.

In the second step of our simulation sequence, we impose the triggering of a detonation at the hottest point and follow it with a grid-based hydrodynamics code[14,15] as it crosses the merged object. The energy release from the nuclear burning in the detonation disrupts the system (see Figure 1). The asymptotic kinetic energy of the ejecta is $1.3 \times 10^{51}$ erg. This is comparable to typical explosion energies of standard SN Ia models[16] arising from Chandrasekhar mass white dwarfs. However, as the total mass of the ejecta is about 1.3 times larger in our model, they have lower velocities on average[17].

Using tracer particles that record the conditions during the explosion phase and a 384-isotope nuclear network[18], we reconstruct the detailed nucleosynthesis of the explosion in the third step. Because of the low densities in the merged object, the nucleosynthesis primarily proceeds in the regime of incomplete silicon burning. Thus, only 0.1 M$_\odot$ of $^{56}$Ni are synthesized and the ejecta consist mostly of intermediate mass elements (1.1 M$_\odot$) and oxygen (0.5 M$_\odot$). Less than 0.1 M$_\odot$ of carbon are left unburned. (For detailed nucleosynthesis results see the Supplementary Information.) Thus, our model fulfils the requirements necessary to reproduce the characteristics of the 1991bg-like class of supernovae.

Finally, we use the structure of the ejecta and the detailed chemical abundances to calculate synthetic light curves and spectra using a Monte Carlo radiative transfer code[19] as required to quantitatively test this model against observations. Owing to the small $^{56}$Ni mass synthesized during the nuclear burning, the synthetic light curves (Figure 2) are faint and decline rapidly compared to those of normal SNe Ia, despite the large total ejecta mass of our simulation (1.8 M$_\odot$). Given that there has been no fine-tuning of the explosion model, the light curves agree remarkably well with those of the 1991bg-like SNe Ia – in both absolute magnitude and colour evolution. Moreover, our model naturally predicts the lack of secondary maxima in the near-infrared (J, H and K) light curves which is a peculiarity of 1991bg-like objects compared to normal SNe Ia.

In detail, however, there are some discrepancies between our model light curves and the observational data. Comparing the difference in brightness at B band maximum and 15 days thereafter we find values between 1.4 and 1.7 – depending on the line-of-sight. This is less than typically observed for 1991bg-like objects (1.9), but at worst similar to the fastest declining normal SNe Ia and substantially faster than for objects which have previously been claimed as possible super-Chandrasekhar explosions (e.g. 0.69 for SN 2006gz[20]). We note that the exact light curve shapes are affected by details of both the nucleosynthesis and the radiative transfer and are thus very sensitive to any systematic shortcomings of the simulations. In particular necessary approximations in the treatment of the ionization state of the ejecta can influence the decline of the light curves[19]. However, as our simulation is only one particular realization of the model, a closer agreement may be found by exploring the initial parameter space.

Figure 3 compares the spectrum of SN 2005bl (a well observed example of 1991bg-like objects) near maximum light to our angle-averaged model spectrum. This illustrates that both the overall flux distribution and the individual spectroscopic features agree remarkably well. Although the features show some variation for different lines-of-sight these are small and the angle-averaged spectrum is representative. For detailed points of comparison particularly relevant to 1991bg-like events, see the Supplementary Information.

The total mass of the system is essentially a free parameter, only a mass ratio close to one is required. Our simulation predicts 1991bg-like events for white dwarf masses of $\sim 0.9$ $M_\odot$. Systems of somewhat heavier white dwarfs will synthesize considerably more $^{56}$Ni owing to their higher densities and will therefore lead to much brighter explosions. Conversely, systems of significantly lower mass will not lead to events classified as SNe Ia. First, they likely fail to trigger detonations because the hot spots occur at too low densities. Moreover, even if a detonation would form, the density of the white dwarf material is too low to synthesize any $^{56}$Ni.

A mass ratio of exactly one is artificial. To be realized in Nature, our model must also permit somewhat smaller values. To verify this, we conducted three additional SPH simulations of the merger phase for binary systems with a primary white dwarf mass of $M_1=0.89 M_\odot$ and secondary masses of $M_2=0.87$, $0.85$ and $0.83 M_\odot$, respectively. All of these systems show evolution similar to our complete simulation. The secondary white dwarf is disrupted and merges violently with the primary and all simulations ignite carbon burning. While the hotspots reach temperatures above $2.5 \times 10^9$ K, the most important difference is that the associated densities are slightly lower. The lowest value (around $3 \times 10^6$ g cm$^{-3}$) is found in the simulation with the least massive secondary. According to the detonation criteria[18], this is still sufficient to ignite a detonation. It is clear, however, that for substantially smaller mass ratios ($M_2/M_1$), the ignition of a detonation will fail as the densities at the hotspots are insufficiently high.

Population synthesis studies[21] predict that a wide range of total masses and mass ratios for merging white dwarf binaries is realized in Nature. For our model of violent mergers with mass ratios close to one, systems with total mass similar to that in our simulation are the most relevant case for SNe Ia. Although lower-mass systems will be more common, they will not lead to a detonation as discussed above. Thus they will not result in SNe Ia, but may be observable as faint short-lived transients. In contrast, more massive mergers will be brighter but much rarer owing to the paucity of high-mass white dwarfs. We used the results of recent population synthesis studies[21] computed with the StarTrack code[22,23] to predict the expected rate of binary mergers suitable for our model relative to those of other possible SN Ia progenitor formation channels. We find that events from our model should occur with a rate of $\approx$ 2-11% of the total SN Ia rate. This fraction is consistent with the observed rate of 1991bg-like supernovae. Observations indicate that subluminous SNe Ia are expected to arise in old (> 1Gyr) stellar populations[24]. While it may seem counterintuitive that the relatively massive (Zero-Age Main Sequence mass of $\sim$ 4-6 $M_\odot$) binaries considered here would produce SNe Ia with long delay times, this is possible if they undergo only one common envelope phase and/or begin their evolution on the Zero-Age Main Sequence with wide orbital separations. In such a case, more luminous SNe would arise from white dwarf mergers originating from more massive progenitors (Zero-Age Main Sequence mass of $\sim$ 6-8 $M_\odot$), or from the single degenerate channel.

The use of SNe Ia to measure the expansion history of the Universe relies on their homogeneity. The possibility of physically different evolutionary paths leading to SNe Ia has therefore been a concern in such studies. We have shown that violent mergers of two massive white dwarfs, even when their total mass exceeds the Chandrasekhar limit, predominantly produce faint events. Thus they are not expected to pollute samples of normal SNe Ia significantly and can be excluded as a source of systematic errors in cosmological distance measurements.

**Acknowledgements** We thank K. Belczynski for supporting the population synthesis analysis and S. Taubenberger and S. Hachinger for helpful discussions. This work was supported by the Transregional Research Centre "The Dark Universe" of the German Research Foundation and by the Excellence Cluster "Origin and Structure of the Universe". The work of F.K.R. is supported through the Emmy Noether Program of the German Research Foundation. We are grateful for computer time provided by the Computer Centre of the Max Planck Society in Garching, where all simulations have been performed.

**Author Contributions** R.P. carried out the hydrodynamical simulations and the nucleosynthesis postprocessing. M.K. performed the radiative transfer simulations. F.K.R. and R.P. worked on development of the hydrocodes. S.A.S. and M. K. developed the radiative transfer code. W. H. and F.K.R. started the project. A.J.R. analysed the results of population synthesis calculations. All authors contributed to the interpretation of the simulations and to the writing of the paper.

**Author information** Correspondence and requests for materials should be addressed to R.P. (rpakmor@mpa-garching.mpg.de).

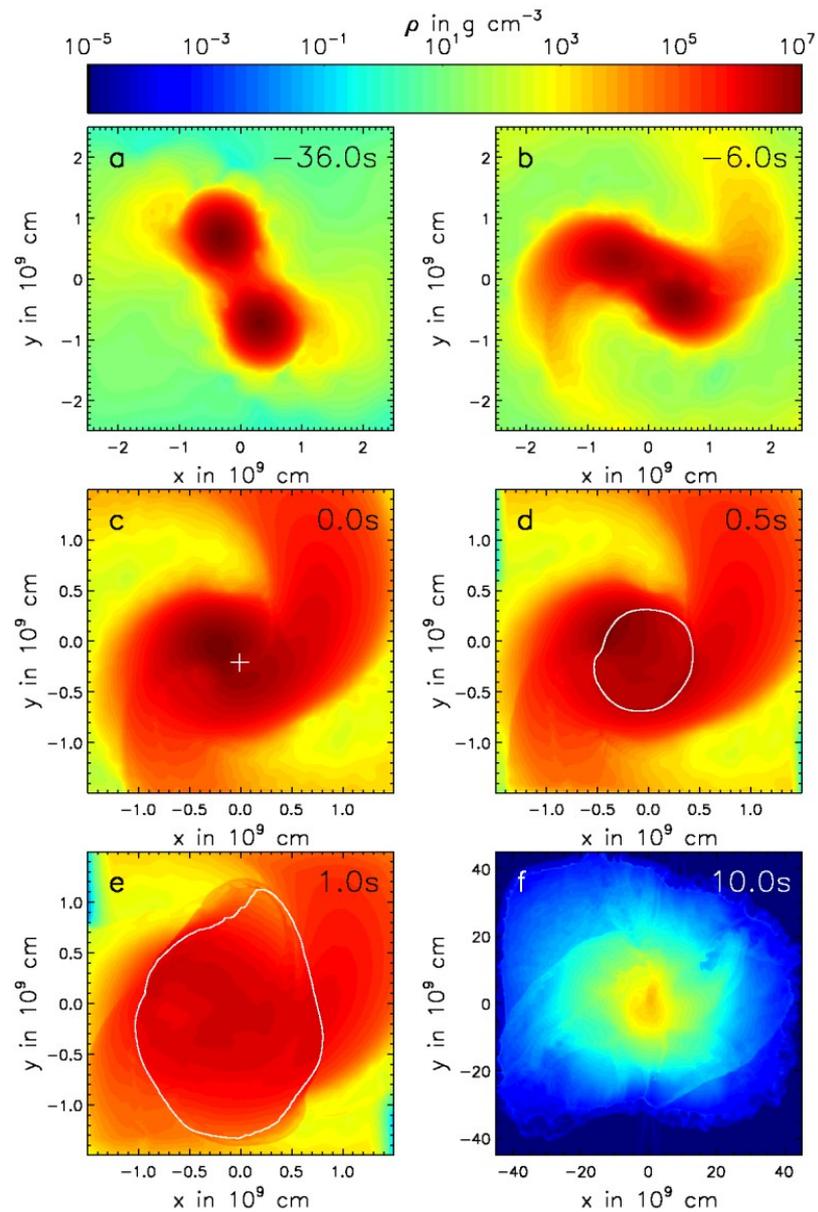

**Figure 1|Time evolution of the binary system.** Slices in the z=0 plane showing colour coded density in logarithmic units. The panels illustrate the complete evolution: starting from a nearly stable orbit (a), dynamic interactions lead to the disruption of one of the white dwarfs (b). In the resulting single object a detonation is triggered (c). The projected location where the detonation starts is marked with a cross. It then propagates through the object (d, e). The white contour shows the position of the detonation shock. Finally the object becomes gravitationally unbound and reaches free expansion (f). The evolution before the detonation occurs is simulated using the smoothed particle hydrodynamics code GADGET2[12] with 2 million particles. At the time the detonation starts (c) the current state of the simulation is mapped onto a uniform Cartesian $512^3$ grid to follow the propagation of the detonation adequately[14,15]. Further detailed information regarding the simulations and the codes can be found in the Supplementary Information.

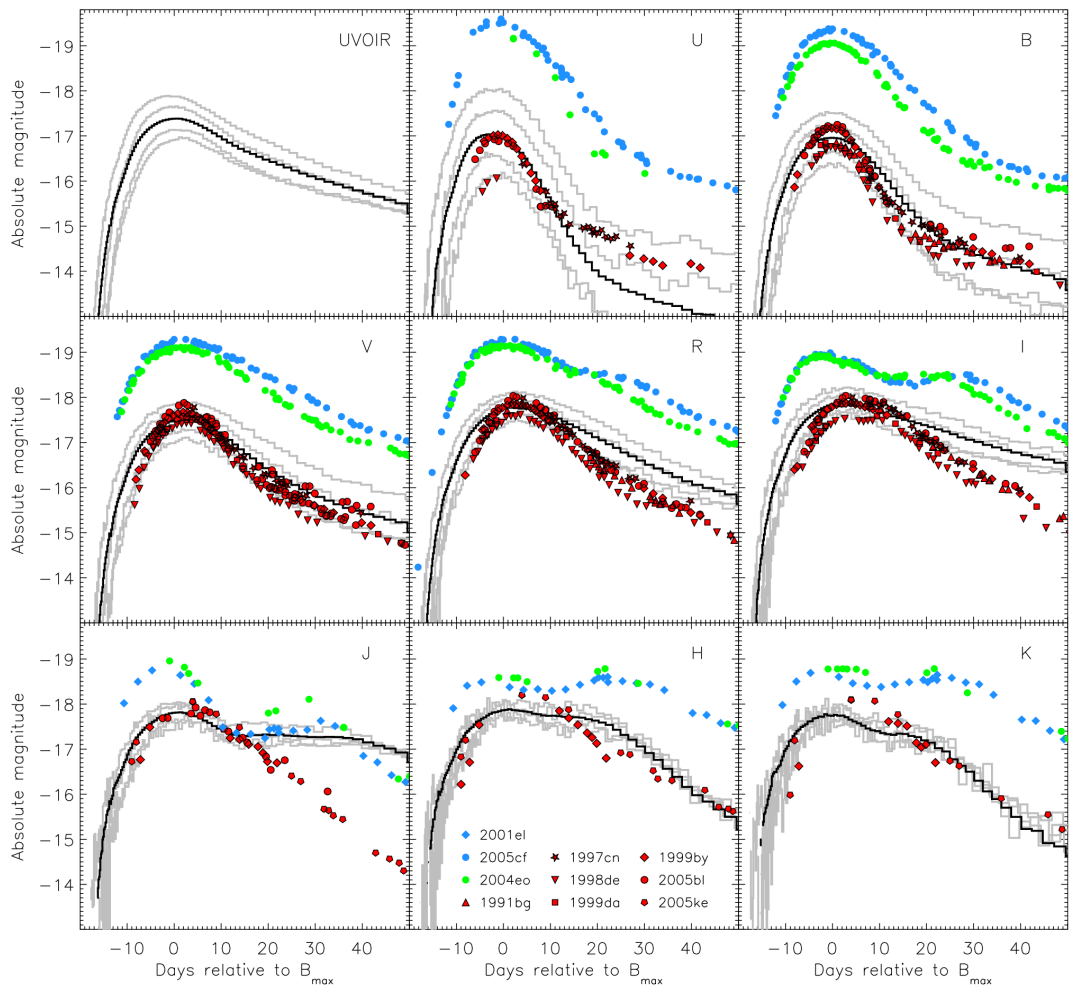

**Figure 2|Synthetic light curves of our model.** From top left to bottom right the histograms show ultraviolet-optical-infrared bolometric (UVOIR) and broad-band U,B,V,R,I,J,H,K synthetic light curves of our model, which have been obtained using the multi-dimensional Monte Carlo radiative transfer code ARTIS[19] after remapping the explosion ejecta to a $50^3$ Cartesian grid. The black histograms show angle-averaged light curves. To indicate the scatter in brightness caused by the model asymmetries, we overplotted four line-of-sight specific light curves (grey histograms). These have been selected from 100 equally sized solid-angle bins such that they represent the full range of the scatter. Time is given relative to B band maximum. The small-scale fluctuations in the histograms are due to Monte Carlo noise in the simulation, which is largest in the NIR bands and at late times in the U band. The region populated by the different lines-of-sight agrees remarkably well with that populated by the sample of observed 1991bg-like SNe Ia shown as red symbols[25]. For SN 1999by, polarization measurements[26] revealed a ~20% degree of asphericity, assuming that the object was observed equator-on. Currently, our radiative transfer simulations do not include polarization, but by extracting the shapes of the surfaces of photon last-scattering we find an upper limit on the asphericity of around 40%. This is consistent with the value obtained for SN 1999by. For comparison the light curves of normal SNe Ia (SN 2005cf[27] and SN 2001el[28]) are also shown. They are much brighter, show a slower decline after maximum light, and distinct secondary maxima in R and redder bands. The same is true for SN 2004eo[29] which represents the faint end of normal SNe Ia.

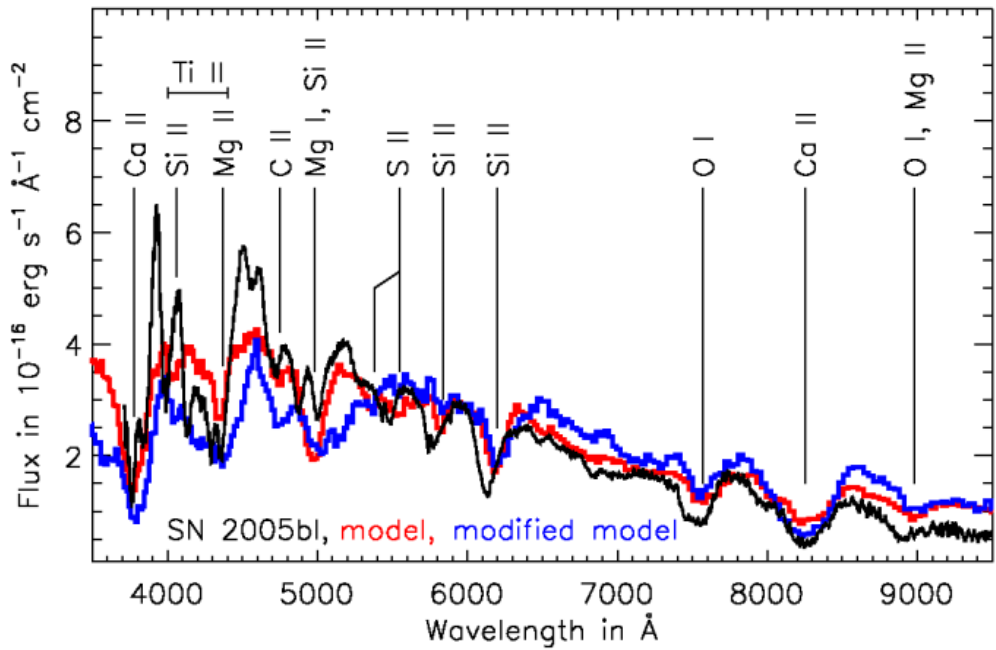

**Figure 3|Synthetic spectrum of our model.** Comparison between SN 2005bl[25] three days before B band maximum (black line) and an angle-averaged synthetic spectrum of our model at the corresponding epoch (red histogram). The strongest features in the model spectrum are labelled. Since the nucleosynthesis yields in the regime of incomplete silicon burning depend strongly on the fuel composition, the detailed spectral features in our model are highly sensitive to the composition of the white dwarfs. This is illustrated by the blue histogram which shows the spectrum obtained for a model adopting different initial composition (see Supplementary Information for details). The effect is strongest in the blue wavelength range, particularly for the Ti II absorption trough between 4000 and 4400 Å.

**Supplementary Information**

Rüdiger Pakmor, Markus Kromer, Friedrich K. Röpke, Stuart A. Sim, Ashley J. Ruiter & Wolfgang Hillebrandt

*This document provides supplementary information for our Letter to Nature. It provides an overview on the different codes, elaborates on our discussion of the simulations, and discusses the comparison of our results with the observational characteristics of 1991bg-like SNe Ia in detail.*

**Smoothed particle hydrodynamics code**

In order to simulate the inspiral of the binary system, the latest version of the Smoothed Particle Hydrodynamics (SPH) code GADGET2[12] is used. As GADGET2 was originally written for cosmological applications, a few modifications are required to extend its applicability to stellar physics. The most important of these are the implementation of a stellar equation of state (EoS) and a nuclear reaction network. For the contributions to the EoS from the electron gas, positrons and radiation, we use tables derived from the Timmes EoS[30]. The ions are treated as a fully ionised ideal gas. A nuclear reaction network comprising 13 isotopes is integrated into the code. This network accounts for **all α-elements up to** $^{56}$Ni ($^{4}$He, $^{12}$C, $^{16}$O, $^{20}$Ne, $^{24}$Mg, $^{28}$Si, $^{32}$S, $^{36}$Ar, $^{40}$Ca, $^{44}$Ti, $^{48}$Cr, $^{52}$Fe, $^{56}$Ni). The nuclear reaction rates are taken from the latest (2009) release of the REACLIB[31] database. The network is solved separately for all active SPH particles which have temperatures in excess of $10^{6}$ K. In an operator splitting approach it is solved separately from the equations of hydrodynamics.

Typically the network requires significantly shorter timesteps; therefore multiple network steps have to be performed for each hydrodynamics timestep. The energy release (or consumption) due to changes in the composition is added to the internal energy of a particle. When combined with the nuclear network it is convenient to solve the EoS with energy as the fundamental quantity rather than entropy (as in the original GADGET2 code). Our EoS is implemented accordingly.

The configuration of the code is very similar to that used in its previous application to impact simulations[32]. All particles have equal mass. The smoothing length is chosen such that each particle has 50 neighbours. The gravitational potential is calculated using the tree method and the gravitational softening length of a particle is set equal to its smoothing length.

**Grid-based hydrodynamics code**

Compared to SPH, grid-based hydrodynamics is superior for following the propagation of shock waves and therefore used in our simulation of the detonation phase[33]. Here, the equations of hydrodynamics are solved according to the Piecewise Parabolic Method[34] while the burning front is modelled with the level-set technique[35]. Originally implemented to follow subsonic deflagration flames[33,35,36], a modification of this scheme[14,15,37] extends it to the tracking of detonations. The burning front is associated with the zero-level set of a signed distance function – positive in the ashes and negative in the fuel – and the fraction of each computational cell swept by this zero-level set is burned into ash with composition depending on the fuel density. The fuel densities in our simulation are too low to reach nuclear statistical equilibrium. Partial silicon burning occurs for fuel densities greater than $\sim 2 \times 10^6$ g cm$^{-3}$. Above $\sim 5 \times 10^4$ g $cm^{-3}$, carbon is converted to oxygen while for lower densities burning ceases. In the simulation, the corresponding difference in binding energies of fuel and ash is released behind the front. Throughout the explosion phase, densities remain low enough that electron captures are negligible (although they are followed in our code). For the grid-based part of the simulation, the same EoS as for the SPH part is adopted. We use a uniform Cartesian grid. The gravitational forces are calculated with a Fast Fourier Transformation (FFT) method. To avoid errors introduced by the periodicity of the FFT, this calculation is performed using a domain twice as large as that for the hydrodynamics. Outside the hydrodynamics domain, the FFT grid is padded with zeros.

The size of the computational domain is set to adequately represent the density distribution while preserving the total mass. With $512^3$ grid cells the optimal choice is a domain size of $3 \times 10^9$ cm. In this case the grid resolution of 60 km is better than the smallest smoothing length of an SPH particle: 110 km at the time of the mapping. This procedure results in a total mass of 1.73 $M_\odot$ on the grid, compared to 1.78 $M_\odot$ in the SPH setup. This small reduction in mass is due to SPH particles lying outside the grid domain. The interpolation onto the grid is performed by assigning to each grid cell the SPH values of density, velocity, internal energy and composition for the centre of the cell. All other quantities are then calculated from the EoS. As the grid code only uses five species (He, C, O and one representative each of the intermediate mass and iron group elements), we must map the 13 species composition to these five species in each cell. The C/O ratio is thereby changed in order to account for burning in the previous SPH part of the simulation.

In order to keep the object in the computational domain while following it to homologous expansion, a moving-grid technique[38] is employed for the grid-based simulation.

**Nucleosynthesis**

To follow the nucleosynthesis in detail, we rely on post-processing[18] of the simulation data. For this purpose, $1.5 \times 10^5$ equal-mass Lagrangian tracer particles are distributed in the object when mapping it onto the grid. These particles are passively advected with the flow and their thermodynamical trajectories are recorded. The trajectories are input to a reaction network calculation which involves 384 nuclei. From this, the detailed composition of the final ejecta is obtained. The initial electron fraction of the unburned material depends on the metallicity of the progenitor stars. We choose this fraction to be 0.5 here. The final yields of the 25 most abundant elements are shown in Table 1. They are consistent with the results of the approximate burning description used in the hydrodynamics code.

**Radiative transfer simulations**

To simulate the radiative transfer we remap the density and composition structure at the end of the hydrodynamics simulation ($t$=100s) to a 3D Cartesian $50^3$ grid. Since the ejecta are free streaming at these times, we assume homologous expansion and let this grid co-expand with the ejecta. Employing the time-dependent three-dimensional Monte Carlo spectrum synthesis code ARTIS[19] we then follow the radiative transfer through the explosion ejecta on this grid.

The total energy available to power the observational display of a SN Ia is determined by the abundances of radioactive nucleosynthesis products. In the simulation, this energy is divided into discrete energy packets, which are initially distributed on the grid according to the distribution of the radioactive nuclides and then follow the homologous expansion until they decay. Upon decay they convert to bundles of γ-ray photons which propagate through the ejecta. ARTIS contains a detailed treatment of γ-ray radiative transfer and accounts for interactions of γ-ray photons with matter by Compton scattering, photo-electric absorption and pair production.

Assuming instantaneous thermalization of absorbed γ-ray photons, the energy is transformed into ultraviolet-optical-infrared photons enforcing statistical and thermal equilibrium. Using a detailed wavelength-dependent opacity treatment, ARTIS solves the radiative transfer problem self-consistently with the ionization and thermal balance equations. Excitation is treated approximately by assuming local thermodynamic equilibrium, which is expected to be a good approximation at least around maximum light. A generalised treatment of line formation[39,40], including about 500,000 individual atomic line transitions[41] in the Sobolev approximation[42], allows for a detailed treatment of radiation-matter interactions including a parameter-free treatment of line fluorescence. This is essential since the opacity in SNe Ia is dominated by the wealth of lines associated with the iron group elements.

The outcome of the radiative transfer simulation with ARTIS is a sequence of synthetic spectra. To obtain synthetic light curves which illustrate the complete evolution of the emitted radiation during the photospheric phase of our model, we integrate these over the appropriate filter functions[43,44].

In the main article, we pointed out that although the general light curve shapes of our model agree quite well with those of the observed sample of 1991bg-like objects, there are some discrepancies especially in the early decline phase after maximum light. Another difference shows up at later times (>20 days post B band maximum). While we continue to obtain reasonable agreement with the observed light curve shapes in the U, B, V, H and K bands until the end of our calculations, discrepancies appear in the I and J band light curves in the sense that our simulations remain too bright. We find that deexcitation of Ca II is responsible for a significant fraction of the flux which emerges in these bands at late times and therefore we suspect that these discrepancies likely have a common origin. This may simply be an artefact of the shortcomings of our radiative transfer simulations for this particular ion or in the details of the nucleosynthesis in the complex regime of incomplete silicon burning.

**Rates and formation channels**

To estimate the event rates expected to arise from our model relative to other potential SNe Ia formation models, we analysed results from population synthesis calculations[21] which were based on the StarTrack code[22,23].

Here, we consider three classes of potential SN Ia progenitors:

(1) all binaries of two CO white dwarfs (double degenerate system; DDS) that merge and have a total mass above the Chandrasekhar mass limit,

(2) all binaries of two CO white dwarfs that merge and have a total mass above the Chandrasekhar mass limit, with a primary mass between 0.85 and 1.05 $M_\odot$ and a mass ratio 0.9<q<1 (these are the progenitors of our model under study),

(3) all CO white dwarf-normal star binaries (single degenerate systems with hydrogen-rich donors; SDS) that may evolve into a SN Ia.

We find that all of the progenitors appropriate for our model (class (2) of the above) originate from binaries with Zero-Age Main Sequence mass ratios > 0.8, and encounter phases of both stable Roche-Lobe overflow as well as unstable mass transfer (a common envelope phase[5,45]). However, the double white dwarfs of class (2) are formed through more than one evolutionary channel, giving rise to double white dwarf mergers with both short (< a few Gyr) and very long (> 8 Gyr) delay times. Below, we briefly outline the evolutionary histories for all double white dwarf mergers originating from class (2) and we divide them into three sub-categories: (2a), (2b) and (2c).

The CO-CO white dwarf mergers with delay times less than 3 Gyr typically originate from binaries with Zero-Age Main Sequence component masses ~ 4.8-7.5 $M_\odot$, and evolve into CO-CO white dwarfs through two different evolutionary channels ((2a) and (2b)), while progenitors with very long (~ 10 Gyr) delay times form from yet a third evolutionary channel (2c), consisting of binaries with initially less massive binaries (component Zero-Age Main Sequence masses ~ 3.8-4.5 $M_\odot$).

We find that the sub-category (2a) progenitors with very prompt delay times (prompt meaning t < 100 Myr[46]) originate from binaries which are generally on close initial orbits (initial separation $a_0$ < 200 $R_\odot$) and additionally have the highest initial Zero-Age Main Sequence component masses ~ 6-7.5 $M_\odot$. However, given their higher initial masses, these progenitors are somewhat rare given the assumed initial mass function[47], and comprise only ~one quarter of the SNe Ia with delay times < 3 Gyr. These more massive binaries evolve very quickly (the primary white dwarf is formed within ~ 50 Myr), and during the evolution, encounter two common envelope phases, effectively bringing the stars on a rather close orbit within a short time. The first common envelope event is encountered between a CO white dwarf and a subgiant star (mass ratio ~ 10), and upon ejection of the common envelope we find a CO white dwarf and a helium-burning star (orbital separation a ~ 1-2 $R_\odot$). As the helium star continues to evolve and expand, it once again fills its Roche Lobe and a second common envelope phase ensues (mass ratio ~ 2), thus bringing the stars on a very close (a < 0.1 $R_\odot$) orbit. Since the timescale associated with orbital angular momentum loss due to gravitational radiation is a strong function of binary separation ($t_{gr} \sim a^4$), the white dwarfs are brought into contact and merge soon after emerging from the second common envelope. Most of these prompt SNe Ia progenitors harbour relatively massive white dwarfs ($\sim 0.9\text{-}1 \, M_\odot$) compared to their larger delay time counter-parts (see below). Given this fact, it is unlikely that this particular formation channel leads to 1991bg-like events, as systems with heavier white dwarfs (both components $> 0.9 \, M_\odot$, as already discussed in the manuscript) will likely lead to more luminous SNe Ia.

By contrast, progenitors of sub-category (2b) with delay times $\sim 1\text{-}3$ Gyr originate from binaries with lower Zero-Age Main Sequence component masses of $\sim 4.8 \text{ - } 5.8 \, M_\odot$ with a wider variety of initial separations ($a_0 \sim 80\text{-}1000 \, R_\odot$), and during their evolution, undergo only one common envelope phase. These binaries encounter a common envelope phase when the primary is a helium-burning star and the mass-losing companion is a red giant; upon ejection of the common envelope both components are helium burning stars on a relatively close orbit ($a \sim 1\text{-}2 \, R_\odot$). However, unlike the (2a) systems with very prompt ($t < 100$ Myr) delay times, once the primary (first-formed helium star) fills its Roche Lobe, mass transfer is stable (mass ratio $< 1$) and thus a second common envelope phase is avoided. Both stars eventually form a CO white dwarf, and the system will take $\sim$ one to a few Gyr to reach contact and merge due to emission of gravitational radiation. If this formation channel leads to SNe Ia, it is unlikely that such SNe will be found in regions of very active star formation such as in young starburst galaxies, though they also will not likely be found within very old elliptical galaxies where star formation has ceased several Gyr ago.

SN Ia progenitors with long delay times ($t > 8$ Gyr to a Hubble time) are formed through yet a third evolutionary channel (2c). These progenitors originate from binaries with an even larger spread in initial separations ($a_0 \sim 200\text{-}2000 \, R_\odot$), and have the smallest Zero-Age Main Sequence component masses ($\sim 3.8\text{-}4.5 \, M_\odot$). For these less massive binaries, the (only) common envelope phase is encountered when the primary has already evolved into a CO white dwarf, and the mass-losing star is a bloated late-AGB star (wide pre-common envelope orbital separation of a > 700 $R_\odot$). The orbital separation upon ejection of the common envelope, which leaves behind two CO white dwarfs, is relatively small (a ~ 3 $R_\odot$), though not sufficiently small so as to enable the stars to reach contact by emission of gravitational radiation within a few Gyr. Thus, SNe Ia arising from this evolutionary channel have very long delay times, and are not expected to occur in young galaxies with active star formation, but are likely to be found among old stellar populations (e.g., old elliptical galaxies).

We find that regardless of the favoured evolutionary model leading to normal SNe Ia (class (1) or class (3)), the progenitors in class (2) will contribute approximately 2-11% of all SNe Ia. As noted in the main article, this is consistent with observations[24] within the uncertainty estimates. If we assume that DDS (class (1)) SNe Ia are the progenitors of normal SNe Ia, then SNe Ia originating from class (2) comprise roughly 2% of the total. If, however, we assume that all normal Type Ia supernovae originate from class (3), and that most CO-CO white dwarf mergers are incapable of leading to a thermonuclear explosion, then SNe Ia originating from class (2) comprise roughly 11% of all SNe Ia. The results from population synthesis calculations also allow us to predict the relative frequency of SNe Ia formed via our model depending on stellar population age (see below).

**Comparison with observations of 1991bg-like SN Ia**

Observationally, 1991bg-like SNe are characterized by distinct features. Here we show that our model can accommodate all of these characteristics leading to the conclusion that it accounts for this class of SNe Ia. We note, however, that since they are observational signatures, not all of them necessarily constrain the explosion physics but some are, from the explosion modelling perspective, details.

*Low luminosity*

We obtain a peak magnitude of ~ -17 mag in B-band, exactly as required for 1991bg-like SNe Ia. This prediction is a direct consequence of the low $^{56}$Ni mass synthesized in our simulations. In principle, more massive mergers could produce more luminous events, but they are expected to be very rare.

*Narrow light curves*

Our model predicts light curves which are narrow compared to those of typical SNe Ia. This is as observed for 1991bg-like SNe Ia. Close comparison shows that our synthetic light curves fade slightly less quickly after maximum than observations of 1991bg-likes suggest. However, there are caveats associated with the radiative transfer calculations which may affect the details here. Using a different approximation for the treatment of ionization produces a change of 0.3 in $\Delta m_{15}(B)$ (the parameter used to quantify the light curve decline rate). This degree of sensitivity is comparable to the difference between our state-of-the-art NLTE treatment (which yields $\Delta m_{15}(B) = 1.4$) and the observed value ($\Delta m_{15} \sim 1.8$).

*Red colours*

The colours of our model are in excellent agreement with the observations of 1991bg-like events. For example, around maximum light our model gives a mean B-V colour of 0.53, V-R of 0.27 and V-I of 0.36. Observed 1991bg-like SNe (SN2005bl, specifically) have B-V, V-R and V-I of 0.61, 0.21 and 0.25, respectively compared to 0.05, 0.06 and -0.16 for a normal SNe Ia (SN2004eo, specifically).

*Preference for an old stellar population*

Recent observations find that rates of broad-lightcurve SNe Ia are proportional to the star formation rate in their environment while rates of narrow-lightcurve SNe Ia are not[24]. In order to test this, we used the population synthesis results described in the section above.

A preference for our model producing events in old stellar populations requires that the ratio of systems in progenitor class (2) to the progenitors of normal (broad lightcurve) SNe is larger for long delay time systems than for short delay time systems. This ratio depends on the assumption of how normal SNe are produced[48]. Arguing that these normal SNe Ia result from single degenerate systems (class (3) of the above), the required trend is predicted by the StarTrack population synthesis code: The ratio of systems in class (2) to systems in class (3) is 0.11 for delay times between 0-3 Gyr, and increases to 0.90 for delay times 10-13 Gyr.

If, instead of single degenerates, all double degenerates (in class (1) above) produce the normal SNe Ia, we require that the ratio of systems in class (2) to systems in class (1) follows the same trend: the ratio of class (2) to class (1) is 0.01 for delay times shorter than 3 Gyr, and increases to 0.15 for delay times between 10-13 Gyr.

Thus, independent of the assumption of the progenitors which lead to the formation of normal SNe Ia, our model predicts a preference for finding the corresponding class (2) events in old stellar populations.

*Low expansion velocities*

In agreement with observations of 1991bg-like SNe, our model predicts expansion velocities (which are measureable via line features in the spectrum) that are significantly low compared to normal SNe Ia. Low velocities are a generic and physical consequence of our model since it involves low explosion energies and high ejecta masses. The exact values, however, depend sensitively on details of the description of nuclear burning in the simulations. Comparing with the observations of SN 2005bl our predictions for the line velocities are slightly too low (e.g. by about 5 per cent for the Ca II NIR triplet at maximum light and 25 per cent for Si II and O I). However, discrepancies at this level are not a major challenge for the model but merely an indication that details of the nuclear burning physics would need to be treated more accurately for perfect agreement.

*Ti II in the spectrum*

Titanium is responsible for a very strong observational feature in the spectra of 1991bg-like objects between 4000 and 4440 Å. However, this feature can be readily formed with only a tiny amount of this element in the SN ejecta. Titanium is synthesised as a trace element in the explosion. As such, the exact amount produced is very sensitive to the chemical composition of the progenitor.

To check this, we computed an explosion identical to that presented in the main article but accounting for solar metallicity and adopting a different initial C/O ratio (consistent with stellar evolution calculations[49]) in the nucleosynthesis calculations. Although this has very little effect on the abundant elements (such as e.g. silicon which changed by only ~13 per cent), the titanium abundance increases by a factor of ~8. This is enough to produce a significant titanium feature in the spectrum as is shown in Figure 3 in the manuscript.

Thus, our model can reproduce the strong titanium feature but it is not a meaningful test of the explosion mechanism – rather it is a powerful probe of the chemical composition of the progenitor.

*Relatively strong intermediate mass elements in the spectrum*

As incomplete silicon burning occurs in most of the material reached by the detonation, intermediate mass elements are synthesised in a large fraction of the ejecta. In agreement with observations of 199bg-like SNe Ia, our radiative transfer calculations show that this leads to strong features in the spectra.

*Strong O I features in the spectrum*

Compared to many models for normal SNe Ia, our model predicts rather large masses of unburnt material (C/O) in the outer ejecta. As can be seen in Figure 3 in the manuscript, this material is responsible for strong O I features in the spectra, as observed for 1991bg-like SNe.

*NIR light curves of SN1991bg-likes do not show distinct secondary maxima*

Normal SNe Ia (see e.g. blue/green lines in Figure 2 in the main article) show two distinct maxima in their near infrared light curves (J, H and K bands). 1991bg-likes do not. This characteristic is reproduced by our model.

**Final yields of the 25 most abundant elements of our model**

| Isotope | Mass | Isotope | Mass |
|---------|------|---------|------|
| $^{28}$Si | 0.56 | $^{29}$Si | $3.0 \times 10^{-4}$ |
| $^{16}$O | 0.49 | $^{31}$P | $2.8 \times 10^{-4}$ |
| $^{32}$S | 0.19 | $^{23}$Na | $2.1 \times 10^{-4}$ |
| $^{24}$Mg | 0.16 | $^{30}$Si | $1.3 \times 10^{-4}$ |
| $^{56}$Ni | 0.12 | $^{33}$S | $1.2 \times 10^{-4}$ |
| $^{12}$C | 0.10 | $^{55}$Co | $5.2 \times 10^{-5}$ |
| $^{20}$Ne | $4.2 \times 10^{-2}$ | $^{53}$Fe | $3.7 \times 10^{-5}$ |
| $^{36}$Ar | $3.1 \times 10^{-2}$ | $^{58}$Ni | $2.9 \times 10^{-5}$ |
| $^{40}$Ca | $2.5 \times 10^{-2}$ | $^{52}$Mn | $2.4 \times 10^{-5}$ |
| $^{52}$Fe | $6.1 \times 10^{-3}$ | $^{53}$Mn | $2.3 \times 10^{-5}$ |
| $^{27}$Al | $7.3 \times 10^{-4}$ | $^{54}$Fe | $1.7 \times 10^{-5}$ |
| $^{57}$Ni | $3.5 \times 10^{-4}$ | $^{44}$Ti | $1.1 \times 10^{-5}$ |
| $^{48}$Cr | $3.4 \times 10^{-4}$ | | |